\documentclass{PoS}
\usepackage{amsfonts, amssymb, amsmath}
\usepackage{style}
\graphicspath{{figs/}}
\newcommand{\FIG}[1]{Fig.~\ref{#1}}

\title{$B_c$ decays from highly improved staggered quarks and NRQCD}

\ShortTitle{$B_c$ decays from HISQ quarks and NRQCD}

\author{Brian Colquhoun\\
KEK Theory Center, High Energy Accelerator Research Organization (KEK)\\ 
Tsukuba 305-0801, Japan
}

\author{
Christine Davies,
Jonna Koponen,
\speaker{Andrew Lytle}
\\
SUPA, School of Physics and Astronomy,
University of Glasgow, Glasgow, G12 8QQ, UK\\
E-mail: \email{andrew.lytle@glasgow.ac.uk}
}

\author{
Craig McNeile\\
Centre for Mathematical Sciences,
Plymouth University, Plymouth, PL4 8AA, UK
}

\author{HPQCD Collaboration\thanks{URL: http://www.physics.gla.ac.uk/HPQCD}}

\abstract{
We calculate semileptonic form factors for the decays 
$B_c \to \eta_c \, l \nu$ and $B_c \to J/\psi \, l \nu$
over the entire $q^2$ range, using 
a highly improved lattice
quark action for charm at several lattice spacings down to $a=0.045$~fm.
We have two ways of treating the $b$ quark:
either with an $\O(\alpha_s)$ improved NRQCD formalism or 
by extrapolating a heavy
mass $m_h$ to $m_b$ in the relativistic formalism. 
Comparison of the two approaches
provides an important cross-check of methodologies in lattice QCD.
Nonperturbative renormalisation of the currents in the relativistic 
theory also allows us then to fix NRQCD-charm normalisation 
for $b$ to $c$ decays such as $B \to D$
and $B \to D^*$.
}

\FullConference{
34th annual International Symposium on Lattice Field Theory\\
24-30 July 2016\\
University of Southampton, UK}

\begin{document}

\section{Introduction}
Many properties of the $B_c$ meson will be measured for the first time at
the LHC. Indeed the LHCb collaboration has already measured the $B_c$ lifetime
in the semileptonic channel 
$B_c \to J/\psi \mu \nu$~\cite{Aaij:2014bva}, with measurements
of $B_c \to J/\psi \tau \nu$ underway. Therefore it is desirable to have
lattice QCD calculations of the form factors parameterising these processes.
A reliable calculation of the ratio $R(B_c \to J/\psi)$ along with experimental
measurements may shed light on existing anomalies in $R(B \to D)$ and
$R(B \to D^*)$~\cite{Lees:2012xj,Huschle:2015rga,Aaij:2015yra,Abdesselam:2016cgx}.

Simulating heavy quarks is a challenge for lattice calculations due to
discretisation artifacts which grow as $(am_h)^n$, where $a$ is the lattice
spacing and $m_h$ is the mass of a heavy quark. By using a highly improved
staggered quark (HISQ) action we are able to simulate charm quarks with
controlled discretisation effects and on the finest available lattices
($a \approx 0.045$~fm) simulate near to the bottom mass. We can also calculate
the form factors with a complementary approach working directly at $m_b$
using an improved non-relativistic (NRQCD) effective theory formalism.
This framework is also being used in other calculations with a $b \to c$
transition in particular $B \to D^*$ and $B \to D$, and others.
A useful output of this calculation will be to assess the systematic
uncertainties present in that framework.

\section{Methodology}
We study form factors for the decays $B_c \to \eta_c \, l \nu$
and $B_c \to J/\psi \, l \nu$.
These can be determined from matrix elements of the $V-A$ operator
between the states of interest. The matrix elements are expressed in
terms of the form factors as
\begin{equation} \label{ps-ps}
\bra{\eta_c(p)} V^{\mu} \ket{B_c(P)} =
f_{+}(q^2) \[P^{\mu} + p^{\mu} - \frac{M^2 - m^2}{q^2} q^{\mu} \] +
f_{0} (q^2) \, \frac{M^2 - m^2}{q^2} \, q^{\mu}
\end{equation}
and
\begin{multline}\label{eq:ps-v}
\bra{J/\psi(p,\ve)} V^{\mu} - A^{\mu} \ket{B_c(P)} = 
\frac{2i \e^{\mu \nu \rho \sigma}}{M + m}
\ve^{*}_{\nu} p_{\rho} P_{\sigma} \, V(q^2) -
(M + m) \ve^{*\mu} \, A_1(q^2) + \\
\frac{\ve^{*} \cdot q}{M + m} \(p + P\)^{\mu} \, A_2(q^2) +
2 m \frac{\ve^{*} \cdot q}{q^2} q^{\mu} \, A_3(q^2) -
2 m \frac{\ve^{*} \cdot q}{q^2} q^{\mu} \, A_0(q^2) \,.
\end{multline}

Here $q=P-p$ is the difference in four-momentum between the $B_c$ and the
outgoing hadron. We work in the frame where the $B_c$ is at rest.
$q^2_{\text{max}}$ corresponds to the outgoing hadron being produced at
rest and $q^2=0$ corresponds to maximum recoil of the outgoing hadron.
The form factors are calculated using two different formalisms, which
are described in more detail below. 

In the fully relativistic calculation,
the vector current normalisation is determined using the relation
\begin{equation*}
\bra{\eta_c(p)} S \ket{B_c(P)} = \frac{M^2-m^2}{m_{b0}-m_{c0}}f_0(q^2) \,,
\end{equation*}
which is absolutely normalised. Comparing this to Eq.~\eqref{ps-ps} at
$q^2_{\text{max}}$, where only the $f_0$ term contributes, determines
the normalisation of $V^\mu$. 
We have normalised the axial vector current using the PCAC relation
\begin{equation*}
p_\mu \bra{0} A^\mu \ket{B_c} = (m_{c0} + m_{b0}) \bra{0} P \ket{B_c} \,,
\end{equation*}
and the form factor $A_1(q^2)$ is extracted from the 
three-point matrix element~\eqref{eq:ps-v}
by arranging that $\ve^* \cdot q = 0$.

We use the highly improved staggered quark 
(HISQ) action~\cite{Follana:2006rc} which systematically removes lattice
artifacts and in particular allows for the simulation of charm quarks with small
discretisation effects~\cite{Donald:2012ga}. 
On the finer sets of ensembles we can simulate 
at masses well beyond $m_c$~\cite{McNeile:2012qf}.
By working in a regime of
$a m_h \lesssim 0.8$ on successively finer lattices, we calculate
the physics of interest over a range in $m_h$ and extrapolate the 
results to $m_b$.

Alternatively we can make use of an improved non-relativistic effective
theory formalism (NRQCD)~\cite{Lepage:1992tx} for which we can work
directly at the $b$ mass. In this case the current operators 
mediating the decay have 
a relativistic expansion, e.g.\ for the spatial axial-vector current
\begin{equation*}
A_k^{\text{nrqcd}} = (1+\a_s z_k^{(0)}+\dotsb)
\[A_k^{(0)} + (1+\a_s z_k^{(1)})A_k^{(1)} + \a_s z_k^{(2)} A_k^{(2)} + \dotsb\] 
\,,
\end{equation*}
where $A_k^{(1)}$, $A_k^{(2)}$, $\dotsc$ are higher order current corrections,
with matrix elements proportional to $1/m_b$. An analagous expression
holds for the vector current.
The matching coefficient $z_k^{(0)}$ has been computed in
QCD perturbation theory~\cite{Monahan:2012dq,Colquhoun:2015oha}, 
and there is a systematic uncertainty from
unknown $\O(\a_s^2)$ terms. This is a leading uncertainty in the 
ongoing calculation of $B \to D^*$~\cite{Harrison}, 
which feature the same $b \to c$ currents. 
A comparison of results from the relativistic extrapolation method,
where the normalisation is simpler and does not rely on perturbation theory,
will give an important handle on this uncertainty.

Calculations are performed on $n_f=2+1+1$ HISQ gauge configurations 
generated by the MILC collaboration~\cite{Bazavov:2010ru},
with lattice spacings $a \approx 0.15, 0.12, 0.09, 0.06$, and $0.045$~fm.
All results presented here are on ensembles with $m_s/m_{ud} = 5$,
i.e.\ unphysically heavy pions. Since the processes studied only have valence
heavy quarks, light quark mass dependence is expected to be small.

\section{Results}

\begin{figure}
\centering
\includegraphics[width=0.7\textwidth]{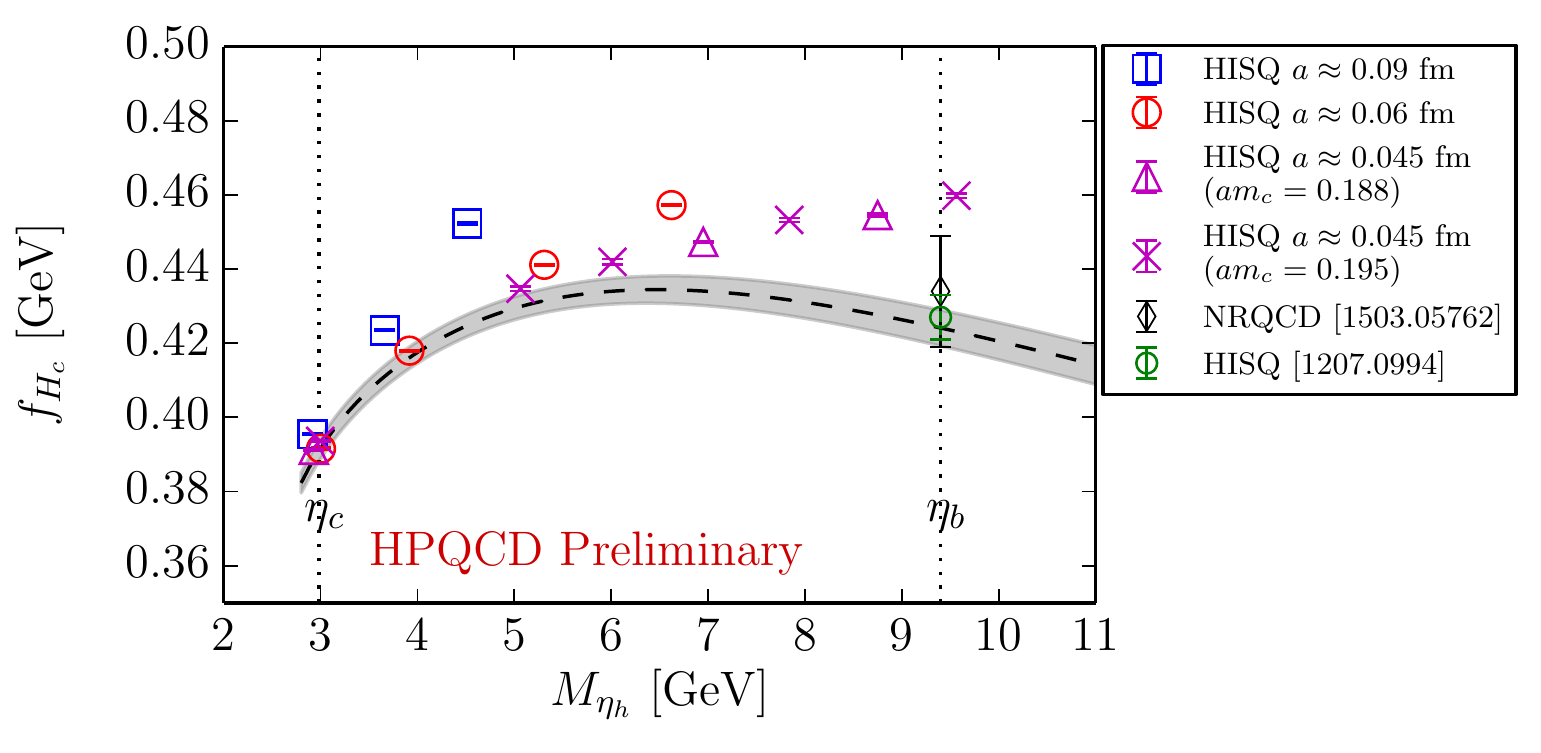}
\caption{
$f_{H_c}$ vs.\ $M_{\eta_h}$ computed on fine, superfine, and ultrafine
ensembles. For each ensemble $m_h$ is varied from $m_c$ to $am_h=0.8$.
The continuum fit result is shown as a grey band.
The results for $f_{B_c}$
are from NRQCD~\cite{Colquhoun:2015oha} and
HISQ on $n_f=2+1$ ensembles~\cite{McNeile:2012qf}.
\label{fHc_vs_Metah}}
\end{figure}

\FIG{fHc_vs_Metah} shows results for the decay constant of a pseudoscalar
meson composed of one charm quark and a heavy quark for 
$m_c < m_h < m_b$ at various lattice spacings. 
For this quantity it is clear that discretisation effects
become sizable as $m_h$ is increased. Nevertheless these effects
remain under control all the way to the $b$ mass on the finest ensemble
with $a \approx 0.045$~fm, and the continuum extrapolation of the combined
data is shown as a grey band in the figure.

\begin{figure}
\includegraphics[width=0.4\textwidth]{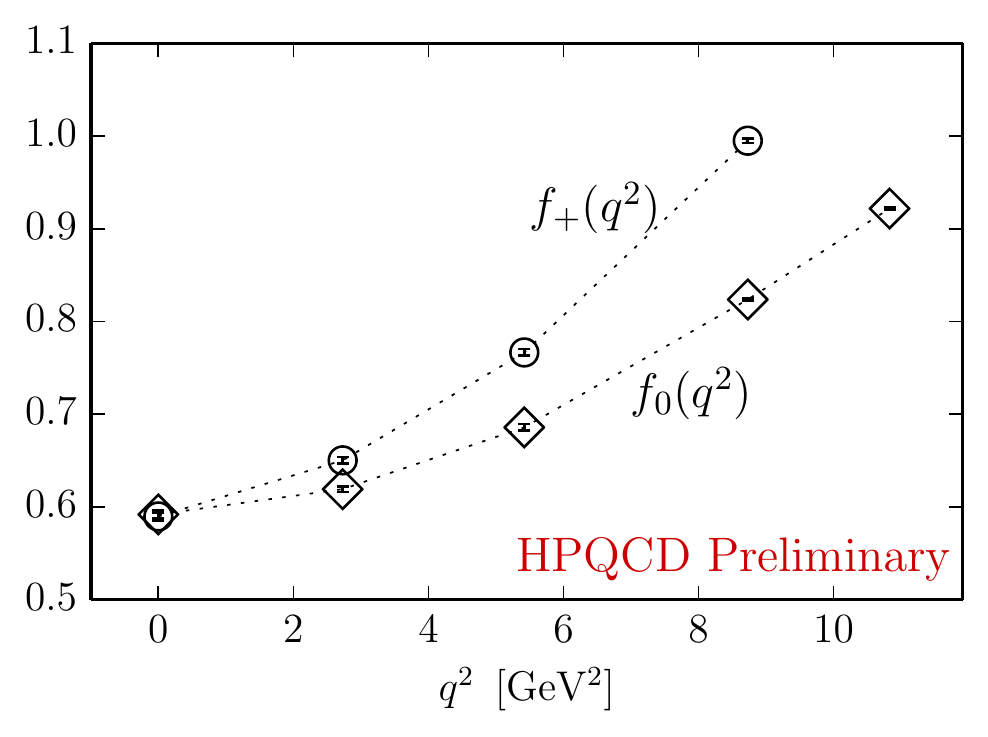}
\includegraphics[width=0.6\textwidth]{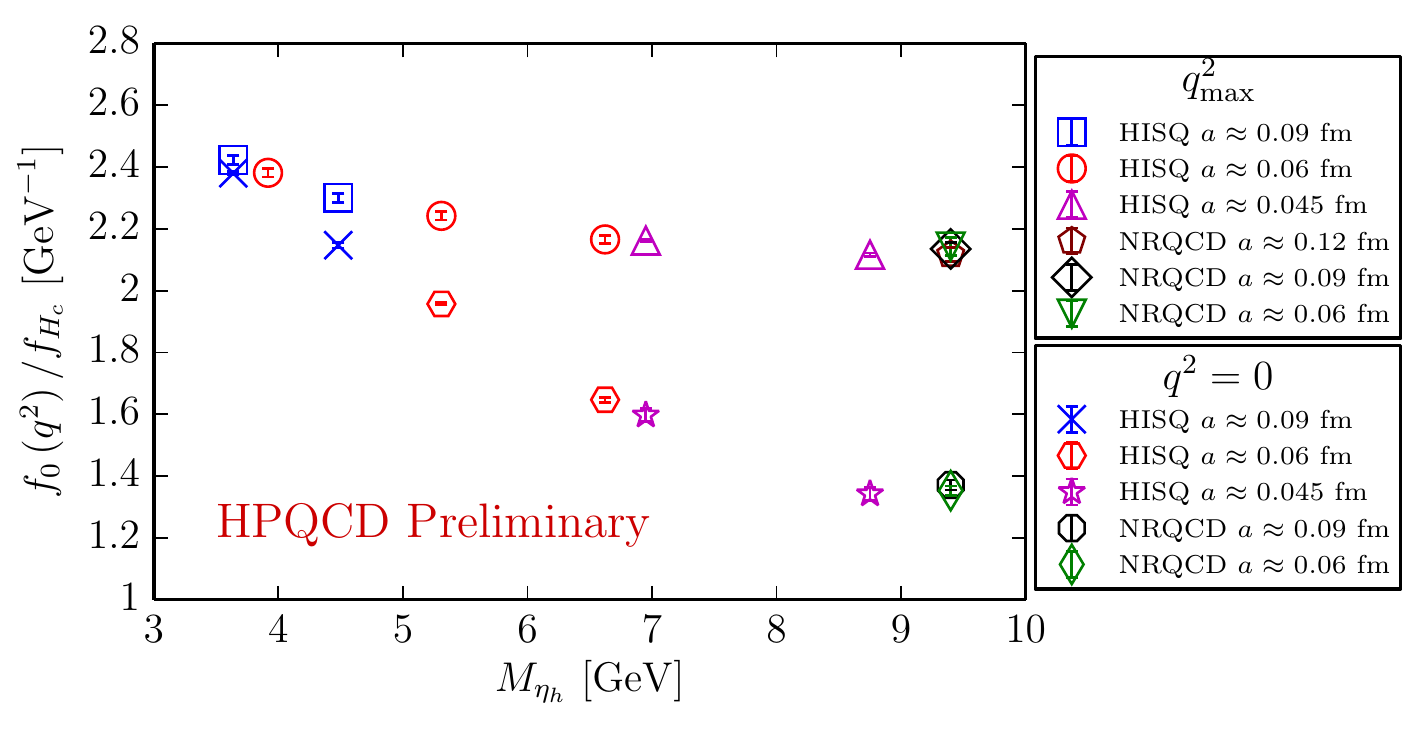}
\caption{
(Left)
$B_c \to \eta_c$ form factors $f_0(q^2)$ and $f_+(q^2)$
calculated using lattice NRQCD, determined over the full $q^2$ range.
(Right)
Results
for $f_0(q^2_{\text{max}})/f_{H_c}$ and $f_0(0)/f_{H_c}$
using the HISQ formalism
as $m_h$ is increased towards $m_b$. The rightmost
points are the corresponding NRQCD results at the physical $b$ mass.
\label{f+-f0}}
\end{figure}

\begin{figure}
\centering
\includegraphics[width=0.5\textwidth]{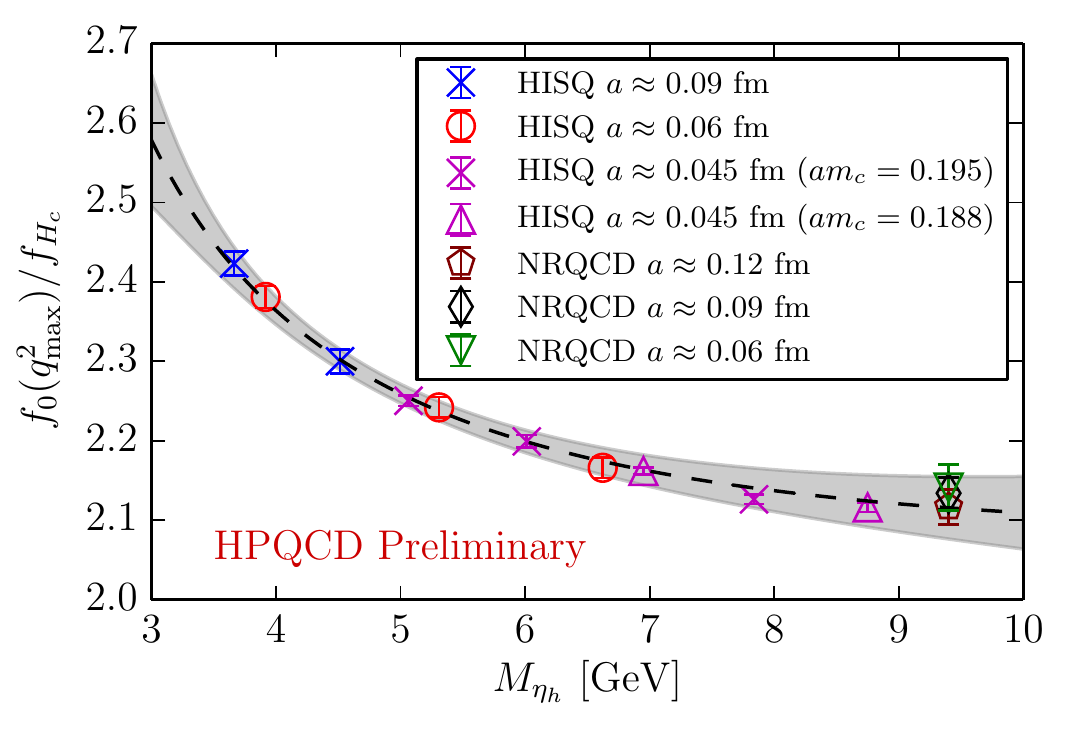}
\caption{
$f_0(q^2_{\text{max}})/f_{H_c}$ vs.\ $M_{\eta_h}$ in the HISQ formalism
on fine, superfine, and ultrafine ensembles. The rightmost points are
NRQCD determinations at the physical $b$ mass.
The grey band shows the HISQ results extrapolated to the continuum.
\label{f0-q2max}}
\end{figure}

\FIG{f+-f0} and \FIG{f0-q2max} show results for the $B_c \to \eta_c$ 
form factors $f_+(q^2)$ and $f_0(q^2)$ 
using both the NRQCD and relativistic formalisms.
\FIG{f+-f0} (left) shows the result for both form factors from NRQCD
calculated on the $a \approx 0.09$~fm ensembler, over the full $q^2$ range.
\FIG{f+-f0} (right) shows the extrapolations in $m_h$
for the points 
$f_0(q^2_{\text{max}})/f_{H_c}$ and $f_0(q^2=0)/f_{H_c}$ 
from the relativistic data 
compared to the NRQCD results from multiple lattice spacings. 
The $f_0(q^2_{\text{max}})/f_{H_c}$ extrapolation
is shown in more detail in~\FIG{f0-q2max} and includes the continuum fit to
the relativistic data.
It is clear from the figure that discretisation effects are small for this 
quantitity throughout the ranges studied, 
and that the continuum result is compatible with the result at $m_b$
coming from NRQCD.

\begin{figure}
\centering
\includegraphics[width=0.5\textwidth]{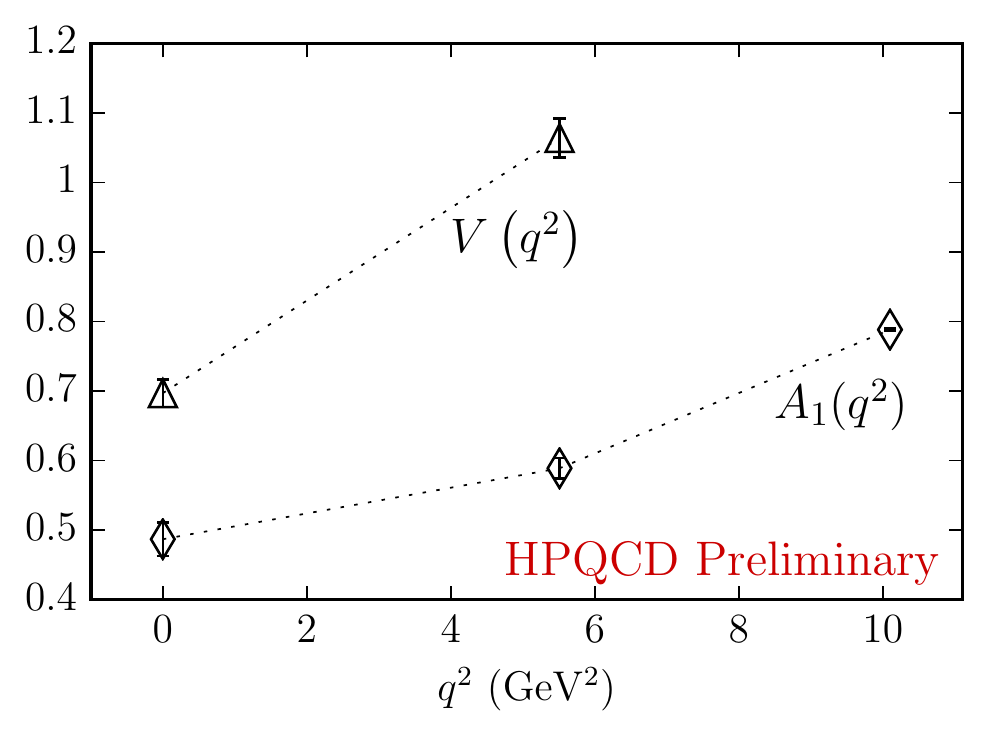}
\caption{$B_c \to J/\psi$ form factors $A_1(q^2)$ and $V(q^2)$
calculated on the $a \approx 0.09$~fm ensemble using lattice NRQCD.
\label{A1-V}}
\end{figure}

\begin{figure}
\includegraphics[width=0.5\textwidth]{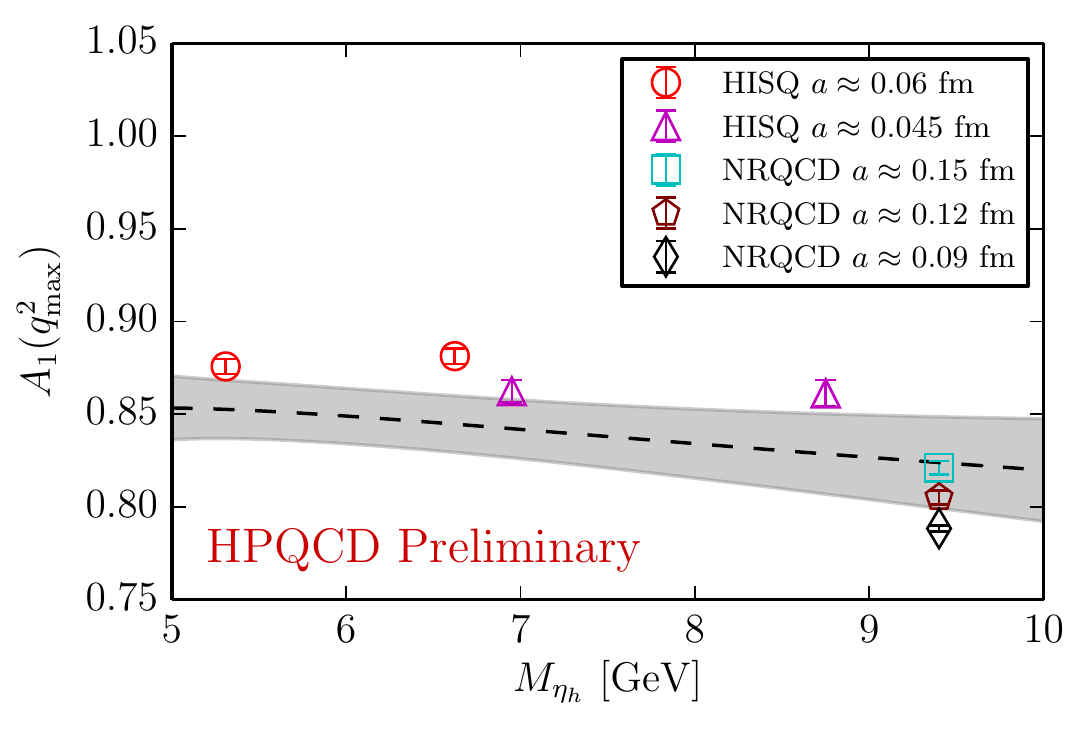}
\includegraphics[width=0.5\textwidth]{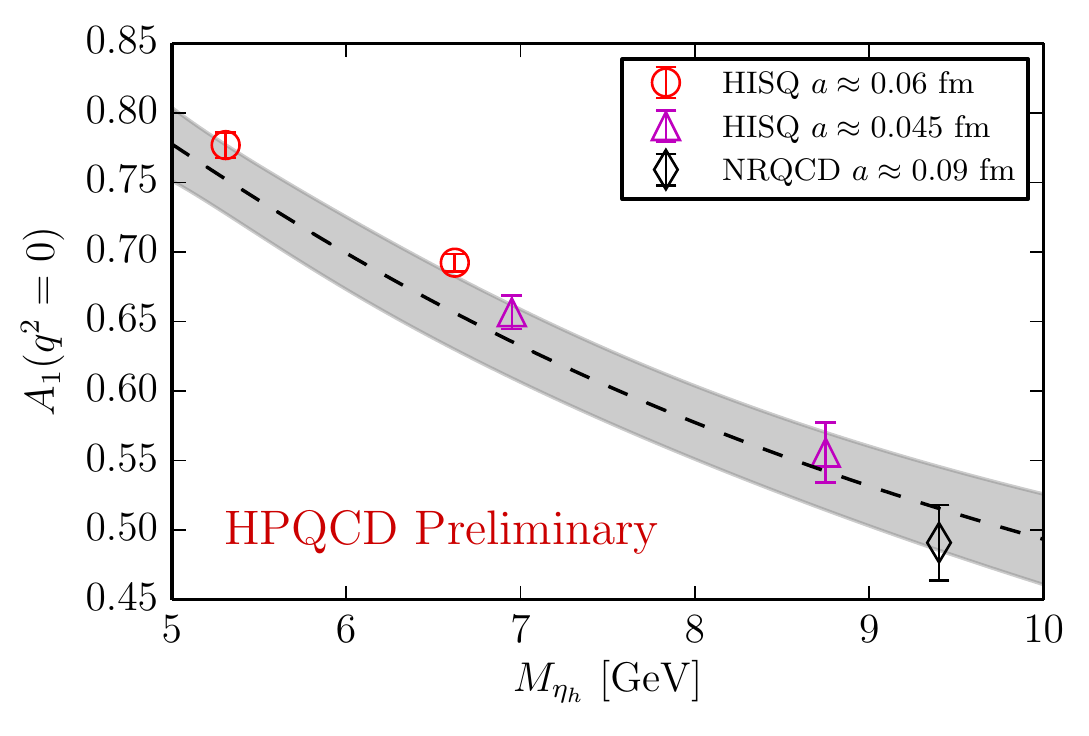}
\caption{
(Left)
Extrapolation in heavy quark mass $m_h$ for the
$B_c \to J/\psi$ form factor
$A_1(q^2_{\text{max}})$ using the relativistic HISQ formalism.
The rightmost points are NRQCD results at the physical $b$ mass,
and the grey band shows the continuum extrapolation of the HISQ results.
(Right)
HISQ results for $A_1(q^2=0)$ vs.\ $M_{\eta_h}$ along with the NRQCD
result computed at the physical $b$ mass. 
The grey band shows the continuum extrapolation of the HISQ results.
\label{A1-q2}}
\end{figure}

\FIG{A1-V} shows NRQCD results for the $B_c \to J/\psi$ form factors
$A_1(q^2)$ and $V(q^2)$ on the $a \approx 0.09$~fm ensemble.
Extrapolations of the relativistic data in $m_h$ are shown 
in \FIG{A1-q2} for the points 
$A_1(q^2_{\text{max}})$~(left) and $A_1(q^2=0)$~(right),
along with the NRQCD results at $m_b$ from multiple lattice
spacings. As in the case of the $B_c \to \eta_c$ form factors,
there is good agreement between NRQCD results and the continuum extrapolations
of the relativistic data.

\section{Conclusions}
We are calculating the form factors for the $B_c$ semileptonic decays 
$B_c \to \eta_c \, l \nu$ and $B_c \to J/\psi \, l \nu$, using two complementary
approaches. One approach utilises the HISQ action on successively finer
lattices to simulate heavy quarks approaching the $b$ mass. The other
works directly at the $b$ mass with improved NRQCD effective theory.
In both approaches we are able to obtain a signal over the full $q^2$ range,
and we see a good agreement between the results of each method.
The NRQCD $b \to c$ currents are also being used in computations 
of $B \to D^*$ and $B \to D$.
Understanding more precisely the normalisations of the 
NRQCD $b \to c$ currents using nonperturbative information provided by
the fully relativistic computation will improve the analyis of this data.

\section{Acknowledgements}
This work was performed on the Darwin supercomputer,
part of STFC's DiRAC facility.


\end{document}